# One-way Hash Function Based on Neural Network

Shiguo Lian, Jinsheng Sun, Zhiquan Wang

Department of Automation, Nanjing University of Science & Technology,
Nanjing, 210094, China, sg_lian@163.com

**Abstract** A hash function is constructed based on a three-layer neural network. The three neuron-layers are used to realize data confusion, diffusion and compression respectively, and the multi-block hash mode is presented to support the plaintext with variable length. Theoretical analysis and experimental results show that this hash function is one-way, with high key sensitivity and plaintext sensitivity, and secure against birthday attacks or meet-in-the-middle attacks. Additionally, the neural network's property makes it practical to realize in a parallel way. These properties make it a suitable choice for data signature or authentication.

**1 Introduction**

Neural networks' confusion and diffusion properties have been used to design encryption algorithms, such as the stream ciphers [1,2] or the block ciphers [3,4]. In fact, neural networks have also a one-way property. For example, if a neuron has multi-inputs and single-output, then it is easy to obtain the output from the inputs but difficult to recover the inputs from the output. These properties make them suitable for hash function [5,6] design. A hash function encodes a plaintext with variable length into a hash value with fixed length, and it is often used in data signature or data authentication. As is known, a secure hash function should satisfy several requirements: one-way, secure against birthday attack and secure against meet-in-the-middle attack. The one-way property makes it impractical to find a plaintext with the required hash value. The hash function should be secure against birthday attack, which makes it difficult to find two plaintexts with the same hash value. It should also be secure against meet-in-the-middle attack, which makes it difficult to find a plaintext whose hash value is the same as one of the given plaintexts. Recently, it was reported that such widely used hash functions as MD5 or SHA-1 are no longer secure. Thus, new hash functions should be studied in order to meet practical applications.

Considering that neural networks have properties suitable for generating hash functions, we try to construct a secure hash function based on a neural network, which not only satisfies the security requirements but also can be efficient-implemented.

## 2 The Proposed Hash Function Based on Neural Network

*The Used Neural Network* In the proposed hash function, the neural network shown in Fig. 1 is used, which is composed of three layers: the input layer, the hidden layer and the output layer. They realize data confusion, diffusion and compression respectively. Let the layer inputs and outputs be $P=[P_0P_1\ldots P_{31}]$, $C=[C_0C_1\ldots C_7]$, $D=[D_0D_1\ldots D_7]$ and $H=[H_0H_1\ldots H_3]$, and the neural network is defined as

$$H = f_2(W_2D+B_2) = f_2(W_2f_1(W_1C+B_1)+B_2) = f_2(W_2f_1(W_1f_0(W_0P+B_0)+B_1)+B_2). \quad (1)$$

where $f_i$, $W_i$ and $B_i$ ($i=0,1,2$) are the transfer function, weight and bias of the i-th neuron layer respectively. Among them, $f_i$ is the piecewise linear chaotic map [7]. It is defined as

$$X(k+1) = f(X(k),Q) = \begin{cases} X(k)/Q, & 0 \leq X(k) < Q \\ (X(k)-Q)/(0.5-Q), & Q \leq X(k) < 0.5 \\ (1-Q-X(k))/(0.5-Q), & 0.5 \leq X(k) < 1-Q \\ (1-X(k))/Q, & 1-Q \leq X(k) \leq 1 \end{cases} \quad (2)$$

where Q is the control parameter and satisfies $0<Q<0.5$. Here, the map is piecewise linear, and it is in chaotic state when $0<Q<0.5$. This chaotic map has some properties suitable for constructing a cipher, such as initial-value sensitivity or parameter sensitivity [7]. If the chaotic map is iterated for T (T is big enough) times, slight difference in the initial-value $X(k)$ or the parameter Q causes large differences in the iterated value $X(k+T)$ [8]. Generally, the chaotic function is iterated for T ($T \geq 50$) times to keep the output's randomness. Based on the chaotic map, the input layer is defined as

$$C = f^T\left(\begin{bmatrix}\sum_{i=0}^{3}w_{0,i}P_i + b_{0,0}\\ \sum_{i=4}^{7}w_{0,i}P_i + b_{0,1}\\ \vdots \\ \sum_{i=28}^{31}w_{0,i}P_i + b_{0,7}\end{bmatrix}, Q_0\right) = \begin{bmatrix}f^T(\sum_{i=0}^{3}w_{0,i}P_i + B_{0,0}, Q_0)\\ f^T(\sum_{i=4}^{7}w_{0,i}P_i + B_{0,1}, Q_0)\\ \vdots \\ f^T(\sum_{i=28}^{31}w_{0,i}P_i + B_{0,7}, Q_0)\end{bmatrix} = \begin{bmatrix}C_0\\ C_1\\ \vdots \\ C_7\end{bmatrix} \quad (3)$$

where $W_0 = [w_{0,0} \; w_{0,1} \; \cdots \; w_{0,31}]$, $B_0$ is 8×1-size, and T is the iteration times (T≥50). Considering that the input of the chaotic map ranges in [0,1], the additions are all module 1. Similarly, the hidden layer and output layer are formulated as follows.

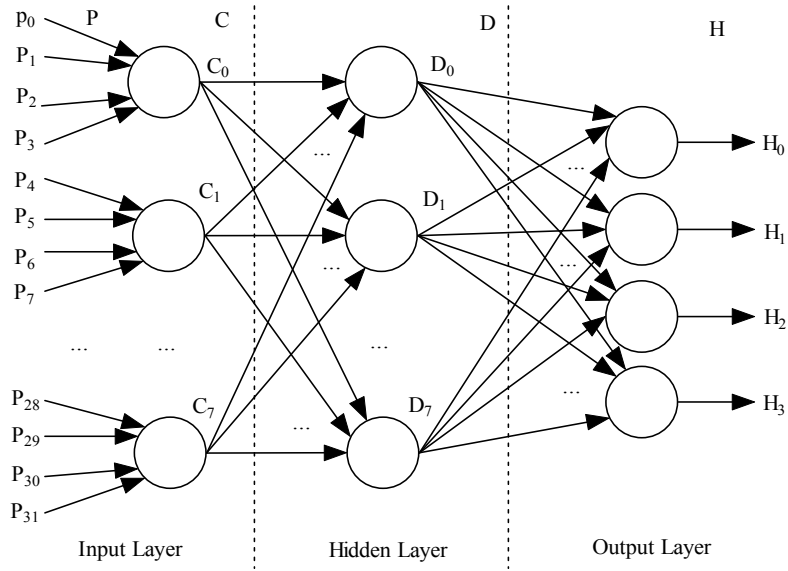

**Fig. 1.** The Three-layer Neural Network

$$D = f_1(W_1C + B_1) = f(W_1C + B_1, Q_1) = \begin{bmatrix}f(\sum_{i=0}^{7}w_{1,0,i}C_i + B_{1,0}, Q_1)\\ f(\sum_{i=0}^{7}w_{1,1,i}C_i + B_{1,1}, Q_1)\\ \vdots \\ f(\sum_{i=0}^{7}w_{1,7,i}C_i + B_{1,7}, Q_1)\end{bmatrix} = \begin{bmatrix}D_0\\ D_1\\ \vdots \\ D_7\end{bmatrix}. \quad (4)$$

$$H = f_2(W_2D + B_2) = f^T(W_2C + B_2, Q_2) = \begin{bmatrix}f^T(\sum_{i=0}^{7}w_{2,0,i}D_i + B_{2,0}, Q_2)\\ f^T(\sum_{i=0}^{7}w_{2,1,i}D_i + B_{2,1}, Q_2)\\ \vdots \\ f^T(\sum_{i=0}^{7}w_{2,3,i}D_i + B_{2,3}, Q_2)\end{bmatrix} = \begin{bmatrix}H_0\\ H_1\\ \vdots \\ H_3\end{bmatrix}. \quad (5)$$

Here, the weight $W_1$ is of 8×8-size, $B_1$ of 8×1-size, $W_2$ of 4×8-size, and $B_2$ of 4×1-size. The hidden layer aims to diffuse the changes in C to the changes in D. The chaotic map f() can be used as the transfer function. In order to keep low cost, the map f() is iterated for only once. The transfer function is $f^T()$ in the output layer where T(T≥50) is the iteration time. The repeated iteration improves the randomness of the relation between H and D, and thus strengthens the cryptosystem.

*The Block Hash* The hash function based on the proposed neural network is shown in Fig. 2, which supports the plain-block with fixed length. That is, the plain-block P composed of 32 data-pixels is encoded into hash value H composed of 4 data-pixels under the control of the user key. Here, each data-pixel is composed of 32 bits, which is quantized (divided by $2^{32}$) to a fractional one ranging in [0,1]. And the result hash bits are extracted from the fractional data-pixels (32-bit from each data-pixel).

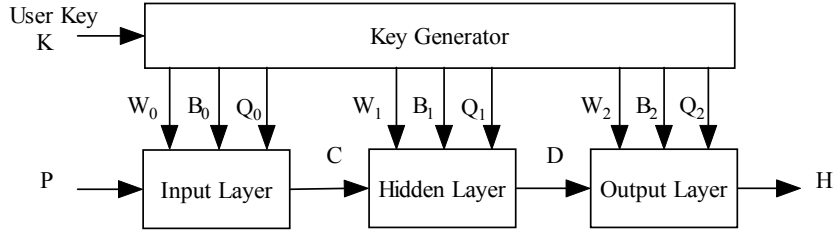

**Fig. 2.** The Proposed Hash Function

Thus, in this condition, the plaintext P consisting of 1024 bits is encoded into the four component hash value H consisting of 128 bits. And the key generator is used to produce the sub-keys: $W_0$, $B_0$, $Q_0$, $W_1$, $B_1$, $Q_1$, $W_2$, $B_2$ and $Q_2$, which is composed of 151 data-pixels. $K=k_0k_1\ldots k_{127}$ is divided into four sub-keys: $K_0=k_0k_1\ldots k_{31}$, $K_1=k_{32}k_{33}\ldots k_{63}$, $K_2=k_{64}k_{65}\ldots k_{95}$ and $K_3=k_{96}k_{97}\ldots k_{127}$. And they are quantized and used to generate all the sub-keys as follows.

$$\begin{cases} X_0(k) = f^{T+k}(K_0, K_1) \\ X_1(k) = f^{T+k}(K_2, K_3) \\ K_s(k) = (X_0(k) + X_1(k)) \bmod 1 \end{cases} \qquad (6)$$

Here, $K_s(k)$ ($k=0,1,\ldots 150$) is the k-th sub-key. The module operation is defined as

$$a \bmod 1 = \begin{cases} a, & 0 \leq a < 1 \\ a-1, & 1 \leq a < 2 \end{cases}.$$

***The Multi-block Hash*** The block hash encodes 1024 bits into 128 bits, and the multi-block hash is proposed to encode the plaintext with binary length into 128 bits. First, the plaintext M is appended to the multiples of 1024. That is, one '1'-bit and some '0'-bits are appended to M. Secondly, it is partitioned into n blocks: $M_0$, $M_1$, … and $M_{n-1}$. Then, these blocks are encoded with the multi-block hash mode shown in Fig. 3. That is, $M_i$'s block hash value $H_{M_i}$ is modulated by its key $K_{M_{i-1}}$. Thus, the final hash value is

$$\begin{aligned} H_M &= K_{M_{n-2}} \oplus H_{M_{n-1}} = (K_{M_{n-3}} \oplus H_{M_{n-2}}) \oplus H_{M_{n-1}} \\ &= \cdots = (K \oplus H_{M_0}) \oplus H_{M_1} \oplus \cdots \oplus H_{M_{n-1}} \end{aligned} \qquad (7)$$

where "$\oplus$" denotes bitwise XOR operation.

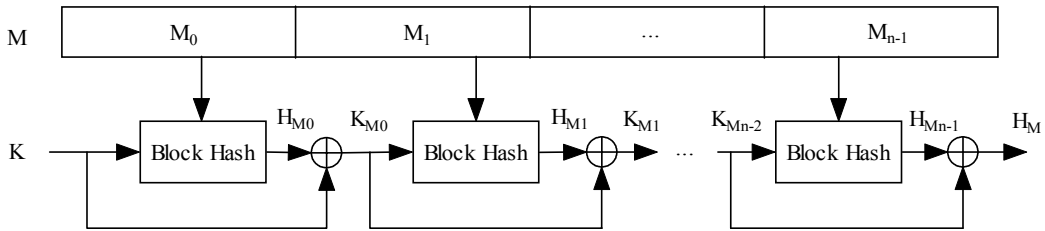

**Fig. 3.** The Multi-block Hash Mode

## 3 Security Analysis

*One-way Property* In the proposed hash function, H is easy to be computed from P and K according to Eq. (1). However, it is difficult to compute P and K if only H is known. In the input neuron layer, the output is

$$C_j = f^T(\sum_{i=4j}^{4j+3} w_{0,i} P_i + B_{0,j}, Q_0) = f^T(Z_j, Q_0) \qquad (8)$$

where j varies from 0 to 7 and $Z_j = \sum_{i=4j}^{4j+3} w_{0,i} P_i + B_{0,j}$.

At first, let's see how to compute $P_i$ from $Z_j$ especially under the condition that $W_{0,i}$ and $B_{0,j}$ are unknown. Two methods can be tried: brute-force attack and select-plaintext attack. For the brute-force attack, 8 data-pixels need at least $2^{256}$ times, which is not practical according to today's computing ability. For the select-plaintext attack, 8 data-pixels need 32 plaintext-key-hash triples. As can be seen, the select-plaintext attack is practical if $Z_j$ is known. However, it is difficult to recover $Z_j$ from $C_j$. According to the chaotic map, it needs $4^T$ ($\geq 2^{100}$) to compute $Z_j$ from $C_j$, which makes it difficult when T (T≥50) is big enough. For the hidden layer and the output layer, the piecewise linear chaotic map is always used, which keeps the two layers one-way.

*High Sensitivity* For a hash function, it is required that different plaintexts or different keys produce different hash values. This property depends on the hash function's plaintext sensitivity and key sensitivity. Experiments are done to test the hash function's plaintext sensitivity and key sensitivity. As an example, M="Cellular neural networks (CNN) chaotic secure communication is a new secure communication scheme based on chaotic synchronization." (ASCII string), and K="0123456789abcdef" (ASCII string). M (1040-bit) is padded by appending a "1" bit and followed by 1007 "0" bits, and thus $H_M$="DF461FA76AC4D5330DF97BD58FC96DAF" (hexadecimal digits). Then, only the first bit of M is changed, and thus $H_M$'="F776C1409C826B7A542FC70965282ED9". Similarly, if only the first bit of K is changed, $H_M$''="1D8168205EEB609AF24903C2519FCAAC". The hamming distance ratio (Hdr) is defined to measure the difference between them, which is the ratio between the hamming distance and 128. Figure 4(a) shows the result of changing each $M_0$-bit, and Figure 4(b) shows the one of changing each K-bit. Seen from the results, all the Hdrs lie near 50%, which means that one bit's change causes a great difference. Thus, the proposed hash function satisfies the sensitivity requirement.

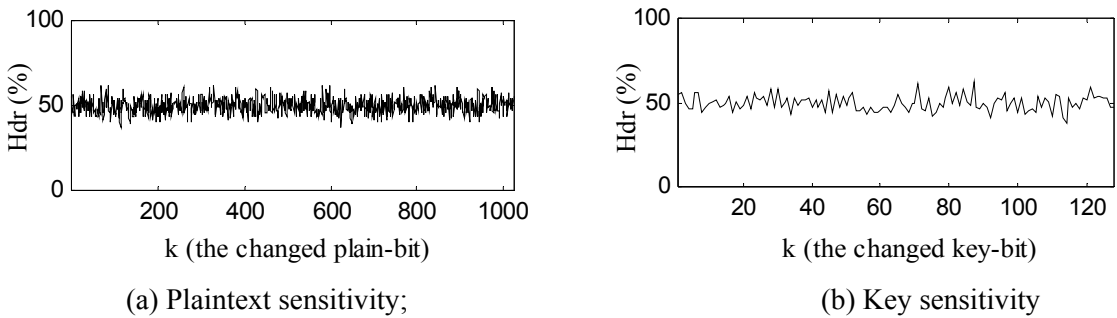

(a) Plaintext sensitivity;  (b) Key sensitivity

**Fig. 4.** Sensitivity Test

***Birthday Attack*** Birthday attack [6] is a typical attack method used to break a hash function. That is, to find a contradiction is similar to find two persons with the same birthday. Thus, for 64-length hash value, the attack difficulty is not $2^{64}$, but much smaller ($2^{32}$). Considering of the practical computing ability, the hash value's length should be at least 128-bit, which keeps the attack difficulty above $2^{64}$. Here, the proposed hash is 128-length, and it is easy to be expanded to 256 or 512. For example, if the output neuron layer's neurons are increased to 8, then the hash value is of 256-bit; if the input plaintext is increased to 2048 and the number of the neurons are doubled, then the hash value is of 512-bit.

***Meet-in-the-Middle Attack*** Meet-in-the-middle attack [6] means to find a contradiction through looking for a suitable substitution of the last plaintext block. If $M=[M_0M_1…M_{n-2}M_{n-1}]$, the expected contradicted one is $M'=[M_0M_1…M_{n-2}M_{n-1}']$. That is, the attack process is just to replace $M_{n-1}$ with $M_{n-1}'$ and keep $H_M$ unchanged, as is shown in Fig. 5. Because $K_{Mn-2}$ is not known, the weight, bias and the chaotic map's parameter are all not known. The attackers may attempt to use many plaintext-key-hash triples, but they cannot obtain $K_{Mn-2}$ because it is in close relation with the key and the previous plaintext blocks. If n=0, there is only one plain-block, which has been analyzed above. Thus, it is difficult to break the hash function with meet-in-the-middle attacks.

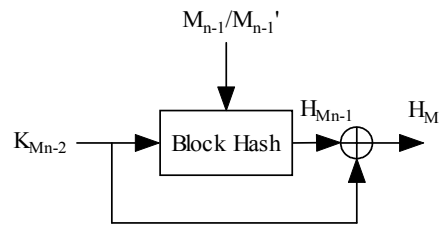

**Fig. 5.** Meet-in-the-Middle Attack

## 4 Computing Complexity

In this hash function, all the three layers and the key generator are realized by multiplication/division and addition/subtraction operations. A neural network's structure makes it practical for parallel realization. Based on this property, the time-efficiency can be improved. The operation numbers of the general-realization and parallel-realization are compared with the ones of the traditional hash functions. Seen from Table 1, the NN-Hash proposed here needs more operations than the traditional ones. However, in a parallel-realization, the operation number decreases greatly, and becomes much smaller than the traditional ones. This property makes it a probable choice for applications with large volumes.

**Table 1.** Comparison of Data Operations
(The plaintext is of 1024 bits, and T=50)

| Operation | Hash Function | | | |
|---|---|---|---|---|
| | MD5 | SHA-1 | NN-Hash | Parallel NN-Hash |
| Multiplication/Division | 296 | 370 | 1088 | 203 |
| Addition/Subtraction | 392 | 330 | 1719 | 291 |

## 5 Conclusions

A secure hash function based on a neural network is presented and analyzed. This hash function adopts the neural network's one-way property, diffusion property and confusion property suitably. The analysis and experiments show that this hash function satisfies the security requirements, and is time-efficient by parallel-realization. Thus, it is proved practical to construct a hash function based on neural networks.

## References

[1] C.-K. Chan and L.M. Cheng. The convergence properties of a clipped Hopfield network and its application in the design of keystream generator, IEEE Transactions on Neural Networks, Vol. 12, No. 2, pp. 340-348, March 2001.
[2] D.A. Karras and V. Zorkadis. On neural network techniques in the secure management of communication systems through improving and quality assessing pseudorandom stream generators. Neural Networks, Vol. 16, No. 5-6, June - July, 2003: 899-905
[3] S.G. Lian, G.R. Chen, A. Cheung, Z.Q. Wang. A Chaotic-Neural-Network-Based Encryption Algorithm for JPEG2000 Encoded Images. In: Processing of 2004 IEEE Symposium on Neural Networks (ISNN2004), Dalian, China, Springer LNCS, 3174 (2004) 627-632.
[4] Liew Pol Yee and L.C. De Silva. Application of multilayer perception networks in symmetric block ciphers. Proceedings of the 2002 International Joint Conference on Neural Networks, Honolulu, HI, USA, Vol. 2, 12-17 May 2002: 1455 – 1458.
[5] Secure Hash Standard. Federal Information Processing Standards Publications (FIPS PUBS) 180-2, 2002.
[6] S.A. Vanstone, A.J. Menezes, P. C. Oorschot. Handbook of Applied Cryptography. CRC Press, 1996.
[7] S. Papadimitriou, T. Bountis, S. Mavroudi, A. Bezerianos. A Probabilistic Symmetric Encryption Scheme for very fast Secure Communication based on Chaotic Systems of Difference Equations. International Journal on Bifurcation & Chaos, Vol. 11, No. 12 (2001) 3107-3115.
[8] S.G. Lian, J.S. Sun, Z.Q. Wang. Security Analysis of A Chaos-based Image Encryption Algorithm. Physica A: Statistical and Theoretical Physics, Vol. 351, No. 2-4, 15 June 2005, Pages 645-661.